\documentclass{PoS}

\title{Differential Reduction Techniques for the Evaluation of Feynman Diagrams}
\ShortTitle{Differential Reduction Techniques for the Evaluation of Feynman Diagrams}
\author{\speaker{S.A.\ Yost~} %\\
        \thanks{This research has been supported in part by US DOE grant DE-SC0000528 and The Citadel Foundation.}\\
Dept. of Physics, The Citadel, 171 Moultrie St., Charleston, SC 29409, USA \\
E-mail: \email{scott.yost@citadel.edu }}
\author{V.V.\ Bytev \\
Joint Institute for Nuclear Research,$141980$ Dubna (Moscow Region), Russia \\
E-mail: \email{bvv@mail.jinr.ru}}

\author{M.Yu.\ Kalmykov %\\
   \thanks{This research has been supported in part by BMBF Grant No.\ 05~HT6GUA, DFG Grants No.\ KN~365/3--1 and KN~365/3--2, and HGF Grant No.\ HA~101.}\\
II. Institut f\"ur Theoretische Physik, Universit\"at Hamburg, 
Luruper Chaussee 149, 22761 Hamburg, Germany \\
Joint Institute for Nuclear Research,$141980$ Dubna (Moscow Region), Russia \\
E-mail: \email{kalmykov.mikhail@gmail.com}}

\author{Bernd A.\ Kniehl${\ }^{\ddag}$ \\
II. Institut f\"ur Theoretische Physik, Universit\"at Hamburg, 
Luruper Chaussee 149, 22761 Hamburg, Germany \\
E-mail: \email{kniehl@mail.desy.de}}

\author{B.F.L.\ Ward %\\
        \thanks{This research has been supported in part by US DOE grant 
		DE-FG02-09ER41600.}\\
Dept. of Physics, Baylor University, One Bear Place, Waco, TX 76798, USA \\
E-mail: \email{BFL{\underline\ }Ward@baylor.edu}}

\abstract{
    Stable reduction methods will be important in the evaluation of high-order 
    perturbative diagrams appearing in QCD and mixed QCD-electroweak radiative
    corrections at the LHC. Differential reduction techniques are 
    useful for relating hypergeometric functions with shifted 
    values of the parameters. We present a proposition
    relating the number of master integrals in the expansion of a 
    Feynman diagram to the number of derivatives in a differential 
    reduction.

}

\FullConference{35th International Conference of High Energy Physics - ICHEP2010,\\
                July 22-28, 2010\\
                Paris France}

\begin{document}

Techniques for the evaluation of physical cross sections have 
traditionally been based on the direct evaluation of Feynman diagrams. 
For obtaining stable numerical results, a deeper understanding of the
analytical structure of Feynman diagrams and Green functions is desirable.
In the framework of dimensional regularization,\cite{dimreg}
any multi-loop Feynman diagram 
can be expanded as a Laurent series in powers of the regularization parameter
$\varepsilon$.  An understanding of the analytic functions in this 
$\varepsilon$-expansion gives useful information about the Feynman diagram.

The hypergeometric function representation is obtained from a multiple 
Mellin-Barnes repre\-sent\-ation\cite{MB} 
of the Feynman integral, which may be written
\vspace{-3mm}
$$
\Phi\left(a_{js},b_{kr},c_i,d_j,\gamma\right) = 
%\frac{1}{(2 \pi i)^m} 
\int_{\gamma+i\mathbb{R}} dz_1 \ldots dz_m
\frac{\Pi_{j=1}^{p} \Gamma\left(\sum_{s=1}^m a_{js} z_s+c_j\right)}
     {\Pi_{k=1}^{q} \Gamma\left(\sum_{r=1}^m b_{kr} z_r+d_k\right)}
z_1^{\alpha_1} \ldots z_m^{\alpha_m} \; , 
$$
where 
$
a_{js}, b_{kr}, c_i, d_j \in \mathbb{R},\ \alpha_k \in \mathbb{C},$ and
$z_k$ are Mandelstam variables.  This integral can be written as a sum of 
multiple residues of the integrated expression, resulting 
in a linear combination of Horn-type hypergeometric 
functions of the form 
\vspace{-3mm}
$$
%\Phi(\vec{\gamma};\vec{\sigma};\vec{x}) 
H_r(\vec{\gamma};\vec{\sigma};\vec{x}) 
= 
\sum_{m_1,m_2,\cdots, m_r=0}^\infty 
\Biggl( 
\frac{
\Pi_{j=1}^K
\Gamma\left( \sum_{a=1}^r \mu_{ja}m_a+\gamma_j \right)
}
{
\Pi_{k=1}^L
\Gamma\left( \sum_{b=1}^r \nu_{kb}m_b+\sigma_k \right)
}
\Biggr) 
x_1^{m_1} \cdots x_r^{m_r} \;,
$$
with
$
\mu_{ab}, \nu_{ab} \in \mathbb{Q},\ 
\gamma_j,\sigma_k \in \mathbb{C}, 
$
and their derivatives with respect to the parameters 
$\sigma_k, \gamma_j$. (See, for example, Ref.\ \cite{Smirnov-DelDuca}.)
In general, the hypergeometric representation of a Feynman diagram
is not unique, since non-linear transformations of the arguments $z_k$ are 
possible.  This is equivalent to quadratic, cubic, and more complicated 
transformations of the associated hypergeometric functions.
A recent example is considered in Ref.\ \cite{Smirnov-DelDuca,KT}. 

It is useful to see what can be learned about Feynman diagrams from
their representation in terms of Horn-type hypergeometric 
functions.  A series of 
publications \cite{Proposition} have presented and explored the following 
proposition:
\begin{itemize}
\item[(i)] Each term in the hypergeometric representation of a Feynman diagram should have the same number of derivatives, independent of the type of hypergeometric functions appearing.
\item[(ii)] 
The number of master integrals required to represent a Feynman diagram should 
coincide with the number of derivatives plus one. 
\end{itemize}

These statements are both based on a \emph{differential reduction algorithm} 
(Ref.\ \cite{reduction} and references therein) which constructs raising and 
lowering operators 
which shift the upper and lower parameters of the hypergeometric function by 
one unit.  The reduction procedure can be used to express the hypergeometric 
functions appearing in the expression for a Feynman diagram in terms of a set 
of basis functions. Specifically, a differential reduction algorithm can be 
applied to any Horn-type hypergeometric function, relating functions with 
shifted arguments according to
\vspace{-4mm}
$$
R(\vec{z}) H(\vec{\beta}+\vec{m};\vec{\lambda}+\vec{n};\vec{z}) 
= 
\sum_{k=0}^L P_k(\vec{z}) \frac{\partial^k}{\partial z_{k_1} \cdots \partial z_{k_r} }
H(\vec{\beta};\vec{\lambda};\vec{z}) \;, 
$$
where $\vec{m},\vec{n}$ are set of integer numbers and 
$R,P_k$ are polynomial functions.  (These relations are also useful in 
constructing the $\varepsilon$-expansion of Hypergeometric 
Functions \cite{expansion1}.)

Consider the standard hypergeometric representation of a Feynman diagram,
\vspace{-3mm}
\begin{equation}
\Phi(n,\vec{j};\vec{z}) = \sum_{a=1}^k S_a(n,\vec{j},\vec{z}) 
H_a(\vec{\beta}_a;\vec{\lambda}_a;\vec{\xi}) \, 
\label{eq1}
\end{equation}
where $\vec{j}$ is a list of the powers of the propagators in the Feynman
diagram, $n$ is the space-time dimension, 
$\vec{\xi}$ are the arguments of the hypergeometric functions, which are 
related the kinematic invariants of the Feynman diagram,
$\{\beta_a,\lambda_a \}$ are linear combinations of $\vec{j}$ and $n$ with 
polynomial coefficients, and $S_a$ are rational functions of the variables 
$\vec{z}$ with coefficients depending on  $n$ and $\vec{j}$.\footnote{In 
general, Eq.~(\ref{eq1}) may contain derivatives of Horn functions with respect
to their parameters. \cite{Smirnov-DelDuca} However, there is some 
evidence \cite{derivatives} that such expressions can again be expressed in
terms of linear combination of Horn-type hypergeometric functions.}
The number of basis elements for the Horn hypergeometric functions $H_a$
can be found by constructing the Pfaff system of differential equations 
for this function.  However, when some of the parameters or differences 
between parameters of the functions $H_a$ are integers, the number of 
derivatives is reduced.  Our proposition (i) is that, regardless of the type 
of functions in the r.h.s. of Eq.~(\ref{eq1}), the number of basis elements 
is the same (up to a module of rational functions). 

Being a sum of holonomic functions, $\Phi(\vec{j};\vec{z})$ is also holonomic.
Thus, the number of basis elements on the r.h.s.\ of Eq.~(\ref{eq1})
is equal to the number of master-integrals $\Phi_k(\vec{z})$ that may be 
derived from the l.h.s. by applying the integration-by-parts (IBP)
technique \cite{ibp}, which may be written symbolically as
$
\Phi(n,\vec{j};\vec{z}) = \sum_{k=1}^h B_k(n,\vec{j};z) \Phi_k(n;z) \;.  
$
The number $h$ of nontrivial master integrals following from IBP 
which are not expressible in terms of gamma functions is then
equal to the number of basis elements $L$ for each term of 
r.h.s. of Eq.~(\ref{eq1}).

The generalized hypergeometric functions of one variable
which have been analyzed in detail in Ref.\ \cite{Proposition}
are all in full agreement with the proposition.

\end{document}